\documentclass[twocolumn,a4,footinbib]{revtex4-1}
\usepackage{relsize}
\usepackage[countmax]{subfloat}
\usepackage{graphicx}
\usepackage{amssymb, amsmath,amssymb,amsfonts}
\usepackage{amsthm,mathrsfs,amsopn}
\usepackage{dcolumn}
\usepackage{bm}
\usepackage{color}
\usepackage[utf8]{inputenc}
\usepackage[table]{xcolor} 
\usepackage{floatrow}
\usepackage{lettrine}

\begin{document}


\title{{Patterns of non-normality in networked systems}}

\author{Riccardo Muolo$^{1}$, Malbor Asllani$^{2,3}$, Duccio Fanelli$^{4}$, Philip K. Maini$^{2}$, Timoteo Carletti$^{5}$, \vspace*{.25cm}}
\affiliation{$^1$Systems Bioinformatics, Vrije Universiteit Amsterdam, De Boelelaan 1108, 1081 HZ, Amsterdam, The Netherlands}
\affiliation{$^2$Mathematical Institute, University of Oxford, Woodstock Rd, OX2 6GG Oxford, UK}
\affiliation{$^3$MACSI, Department of Mathematics and Statistics, University of Limerick, Limerick V94 T9PX, Ireland}
\affiliation{$^4$Dipartimento di Fisica e Astronomia, Universit\'a di Firenze, INFN and CSDC, Via Sansone 1, 50019 Sesto Fiorentino, Firenze, Italy}
\affiliation{$^5$Department of Mathematics and naXys, Namur Institute for Complex Systems, University of Namur,
rempart de la Vierge 8, B 5000 Namur, Belgium}

\date{\today}

\begin{abstract}
Several mechanisms have been proposed to explain the spontaneous generation of self-organized patterns, hypothesised to play a role in the formation of many of the magnificent patterns observed in Nature. 
In several cases of interest, the system under scrutiny displays a homogeneous equilibrium, which is destabilized via a symmetry breaking instability which reflects the specificity of the problem being inspected. 
The Turing instability is among the most celebrated paradigms for pattern formation. {In its original form, the diffusion constants of the two mobile species need to be quite different from each other for the instability to develop. Unfortunately, this condition limits the applicability of the theory.} To overcome this impediment, and with the ambitious long term goal to eventually reconcile theory and experiments, we here propose an alternative mechanism for promoting the onset of patterns. To this end a multi-species reaction-diffusion system is studied on a discrete, network-like support: the instability is triggered by the 
non-normality of the embedding network. The non-normal character of the dynamics instigates a short time amplification of the imposed perturbation, thus making the system unstable for a choice of parameters that would yield stability under the conventional scenario. Importantly, non-normal networks are pervasively found, as we shall here briefly review. 
\end{abstract}

\maketitle

\section{Introduction}
We are surrounded by spatially heterogeneous patterns~\cite{prigogine,murray}. In many applications, the interplay between microscopic entities is modelled by means of reaction-diffusion equations, partial differential equations which govern the deterministic evolution of the concentrations in a multi-species model, across time and space. Homogeneous equilibria of a generic reaction-diffusion system may undergo a symmetry breaking instability, when exposed to a heterogeneous perturbation, and this underlies the self-organization theory of pattern-formation.    

The instability can develop according to different modalities, depending on the specificity of the system under consideration. Among existing approaches, the celebrated Turing instability occupies a prominent role~\cite{turing}. The condition for the emergence of Turing patterns has been elegantly grounded in an interplay between slow diffusing activators and fast diffusing inhibitors~\cite{meinhardt}. {Indeed, the parameter space for which the instability materializes shrinks to zero when the ratio of the diffusion constants tends to $1$, an observation that limits the range of applicability of the theory, at least in its original conception}. To state it differently, the condition for the Turing instability are  solely met in a small region of the available parameter space, {in stark contrast with the diversity of patterns that are routinely found across different fields and scales~\cite{ball}. Considering more than two families of interacting entities and accounting for the role of endogenous or exogenous noise enhance the robustness of the scheme and partially overcome the above limitations \cite{stavans, stavans1}}.

In their pioneering paper, Othmer \& Scriven~\cite{OthmerScriven}, extended the Turing theory of pattern formation to reaction-diffusion systems defined on several discrete lattices, in arbitrary dimensions. More recently Nakao \& Mikhailov~\cite{nakao} extended the analysis to systems defined on complex networks, a setting that is relevant too; for example, multispecies systems confined in compartimentalized geometries~\cite{econet},  optically controlled bioreactors~\cite{bioreactor}, bistable chemical networks~\cite{nikos} and neuroscience~\cite{neuro_turing}. The dispersion relation, that ultimately signals the onset of the instability,  is a function of the discrete spectrum of the Laplacian matrix, i.e. the diffusion operator, associated to the underlying network. Laplacian eigenvalues set  the spatial characteristics of the emerging patterns, when the system under scrutiny is rooted on a heterogeneous support. Patterns for systems evolved on complex graphs  yield a macroscopic segregation into activator-rich and activator-poor nodes~\cite{nakao}. They are termed Turing patterns, because of the analogy with their continuum counterparts.  For undirected (symmetric) networks, the topology defines the relevant directions for the spreading of the perturbation, but cannot impact on the onset of the instability. Network asymmetry can however trigger the system unstable, seeding the emergence of a generalised class of topology driven patterns, which extends beyond the conventional Turing scenario~\cite{asllani}. 

\begin{figure}[!ht]
\centering
\includegraphics[scale=0.23]{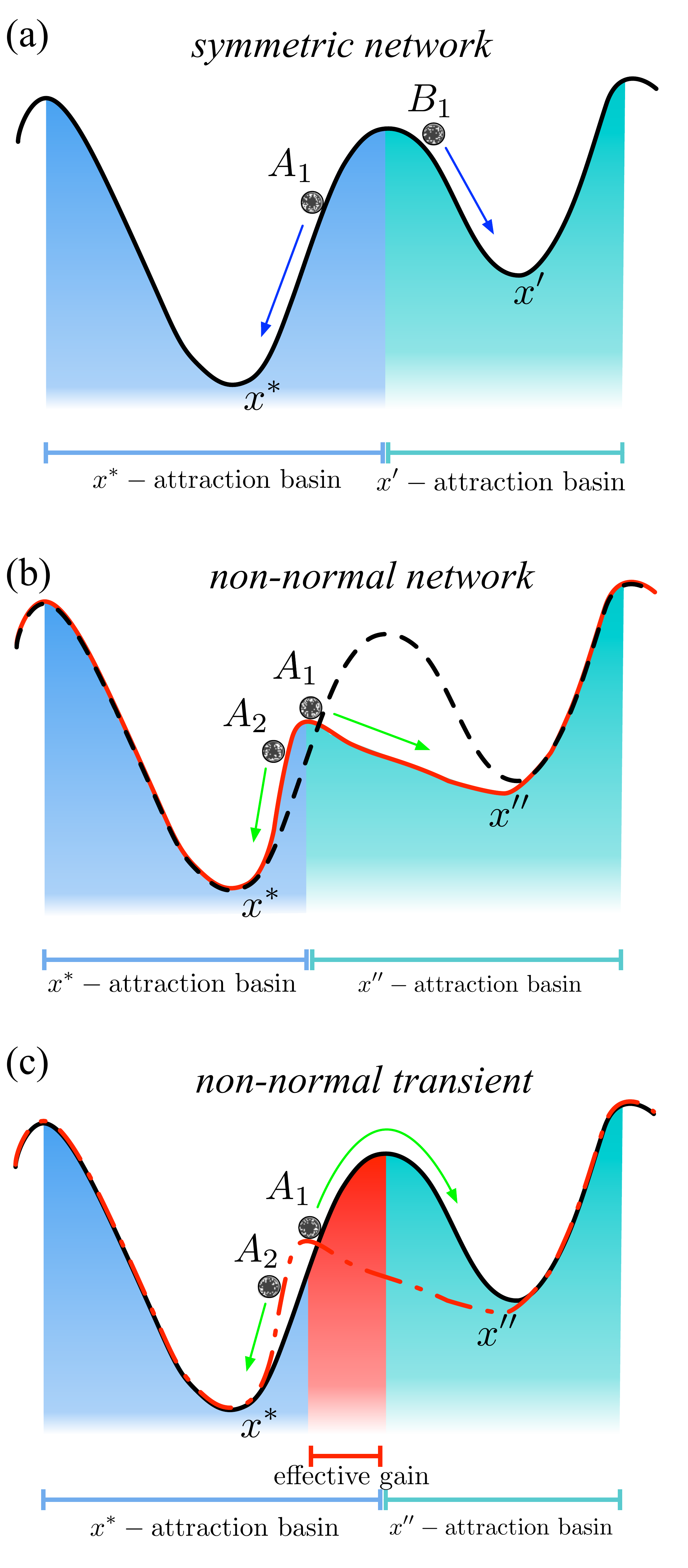}
\caption{\textbf{Attraction landscape.} A schematic layout to depict the landscape of a generic reaction-diffusion system, defined on a symmetric network (a): the basin of attraction associated with the homogeneous equilibrium, $x^*$ (light blue)  is displayed; the latter extends considerably to eventually entwine a large fraction of orbits (as e.g. $A_1$). Only trajectories which are distant enough from the homogeneous fixed point (as e.g. $B_1$) can evolve towards a different, possibly non homogeneous, equilibrium $x^\prime$ (turquoise). Once the dynamics is made to flow on a non-normal support (panel b), the effective basin of attraction of the homogeneous fixed point shrinks considerably. This is the direct signature of the short time amplification of the imposed perturbation, as stimulated by the non-normal character of the underlying support. The amplification makes it possible for the system to overcome the barrier as displayed in (c)  and eventually results in the dynamical landscape, that is pictorially exemplified in (b). }
\label{fig:stabland}
\end{figure}

Starting from these premises, we here set to explore a new situation that is faced when the network of connections is non-normal~\cite{trefethen,nonnormal}, hence inherently asymmetric, but the ensuing dynamics proves stable under a standard linear stability analysis. Otherwise stated, we will deal with homogeneous equilibria which are deemed stable and rely on the short time amplification of the perturbation, as instigated by the non-normal character of the network, to find a path towards the heterogeneous attractor. {Non-normal patterns display a characteristic amplitude, comparable to that associated with their conventional homologues. Small initial perturbations, self-consistently amplified by the non-normality, suffice to cross the barrier which separates the basins of attraction of, respectively, the homogeneous and heterogeneous solutions. The latter can be ideally pictured as the minima of a generalised energy landscape, as schematically highlighted in Fig.~\ref{fig:stabland}. Equally small perturbations cannot trigger the instability when the system is instead defined on a symmetrised version of the non-normal support. The net effect is a contraction of the basin of attraction of the homogeneous fixed point, as produced by the imposed degree of non-normality.}

The role of non-normality has been already studied in~\cite{neubertmurray}, where it was shown that reactivity, i.e. the amplification of the disturbance as seeded by the non-normal nature of the reaction terms, is a necessary condition for the onset of {classical} Turing patterns,  i.e. {patterns ensuing from} a reaction-diffusion system defined on {continuous}. Endogenous noise, or perpetual exogenous perturbations, can effectively contribute to the stabilisation of the patterns~\cite{duccio}, the more effectively the larger the degree of non-normality of the reaction component~\cite{Tommaso}. At variance with the former studies, we shall hereafter focus on the non-normality as stemming from the couplings which define the {spatial support} of the investigated model. The reason for such a choice is twofold: on the one hand, existing reaction models do not exhibit a large non-normal behaviour~\cite{polito}, and on the other hand, a strong non-normality manifests as a natural property of empirical networks, as it has been recently reported in~\cite{nonnormalnet}. 
{In conclusion, based on the above, we can affirm that the non-normal nature of the imposed spatial couplings contributes to significantly enlarging the parameter space where patterns are predicted to occur, potentially increasing the applicability of the theory.}  

The paper is organised as follows: in the next section we will introduce the mathematical framework and then turn, in Section III, to discussing the emergence of pattern for a reaction-diffusion model defined on a non-normal, network-like support. In Section IV we will develop the concept of pseudo-dispersion relation for predicting the onset of the non-normal instability. This serves as a powerful diagnostic tool beyond the linear order of approximation, as routinely employed. Finally, in Section V we summarise up and draw our conclusion. 

\begin{figure}[!ht]
 \includegraphics[scale=.3]{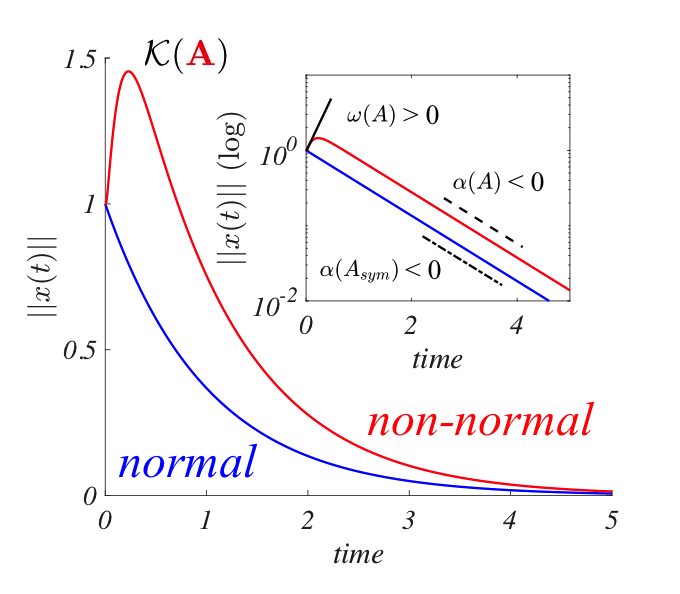}
\caption{{\textbf{Non-normal dynamics.}} Time evolution for the norm of $\textbf{x}$, solution of the stable linear system $\dot{\textbf{x}} = \textbf{Ax}$ for a normal (blue) and non-normal (red) system. In both cases the system is stable, meaning that the spectral abscissa is negative $\alpha(\textbf{A})\equiv\sup\Re\sigma(\textbf{A}) < 0$. The norm $\parallel \textbf{x}(t)\parallel$ can however undergo a short time amplification if the numerical abscissa is positive, $\omega(\textbf{A})\equiv\sup\sigma(\textbf{H})>0$ where $\textbf{H}=\left(\textbf{A}+\textbf{A}^*\right)/2$ is the Hermitian part of $\textbf{A}$.
The norm of the solution is lower bounded by the Kreiss constant $\mathcal{K}(\textbf{A})\leq\sup_{t\geq 0}\parallel \textbf{x}(t)\parallel/\parallel \textbf{x}(0)\parallel$~\cite{trefethen}. } 
\label{tab:table}
\end{figure}
\vspace*{.5cm}

\section{Reaction-diffusion models on networks}

We consider the coupled evolution of two species which are bound to diffuse on a directed network. We introduce the index  $i=1,..., N$ to identify the nodes of the collection and denote by $u_i$ and $v_i$ the concentration 
of the species on the $i$-th node. Local, reactive dynamics is assumed to be governed by the non-linear functions $f(u_i,v_i)$ and $g(u_i,v_i)$. The underlying network is characterised by a binary adjacency matrix $\mathbf{A}$: 
$A_{ij}$ is equal to unity if there is a link that goes from node $j$ to node $i$, and zero otherwise. The species relocate across the network, following available edges and subject to standard Fickean diffusion. For example, relative to species $u$, the net flux at node $i$ is given by {$D_u\sum_{j=1}^N A_{ij}(u_j-u_i)$} where the sum is restricted to the subset of nodes $j$ for which $A_{ij}\neq 0$ and $D_u$ is the diffusion coefficient for species $u$. Collecting this information together, the system under examination obeys the following set of $2N$ ordinary differential equations:
\begin{equation}
\begin{cases}
\dot{u}_i&=f(u_i,v_i)+D_u \sum_{j=1}^N L_{ij}u_j\vspace*{.2cm}\\
\dot{v}_i&=g(u_i,v_i)+D_v \sum_{j=1}^N L_{ij}v_j\, ,
\end{cases}
\label{eq:RD} 
\end{equation}
where the ``dot'' denotes $\dfrac{d}{dt}$, $L_{ij}=A_{ij}-\delta_{ij}k^{in}_i$ are the elements of the Laplacian matrix $\mathbf{L}$ and $k_i^{in}=\sum_jA_{ij}$ stands for the incoming degree of node $i$, i.e. the number of connections pointing towards $i$.  It is worth recalling that standard reaction-diffusion systems on continuous domains are customarily described by a closed set of partial differential equations (PDEs), for concentrations. When discretising the PDE over a finite spatial mesh, the problem can be equivalently cast in the form (\ref{eq:RD}), for a specific choice of the operator  $\mathbf{L}$ and by properly rescaling the coupling constants, so as to account for the selected mesh size, e.g. the $2D$ lattice case.  

As a prerequisite for the forthcoming analysis, we shall assume the existence of a homogeneous equilibrium. This is labelled  $(u^*,v^*)$, and satisfies the condition $f(u^*,v^*)=g(u^*,v^*)=0$. The solution $(u_i,v_i)=(u^*,v^*)$ for all $i$ is further assumed to be stable under homogeneous perturbations. As first intuited by Turing, the aforementioned equilibrium can become unstable upon injection of a small heterogeneous perturbation, which activates the diffusive couplings. These latter are otherwise silenced, as long as the solution stays homogeneous.  Under specific conditions, non-homogeneous disturbances amplify: the system is consequently driven towards different asymptotic attractors, termed in the literature as Turing patterns. To isolate the conditions that make the instability possible, one proceeds with a conventional linear stability analysis. The key idea is to introduce a small perturbation of the homogeneous equilibrium, $\mathbf{x}=(u_1,\dots,u_N,v_1,\dots,v_N)^T-(u^*,v^*)^T$ and insert it in the governing Eq.~\eqref{eq:RD}. By expanding at the first oder in the perturbation, we obtain the following linear system:
\begin{equation}
\frac{d}{dt}\mathbf{x}=\left( \mathcal{J}_0+\mathcal{L} \right) \mathbf{x}\, ,
\label{eq:RDLin} 
\end{equation}
where the Jacobian matrix $\mathcal{J}_0$ is
\begin{equation}
\mathcal{J}_0=\left(\begin{matrix}f_u \mathbf{1}_N & f_v \mathbf{1}_N\\ g_u \mathbf{1}_N &g_v \mathbf{1}_N\end{matrix}\right)=\mathbf{J}_0\otimes \mathbf{1}_N\, ,
\label{eq:J0}
\end{equation}
for $\mathbf{J}_0=\left(\begin{matrix}f_u  & f_v \\ g_u &g_v \end{matrix}\right)$ the $2\times 2$ Jacobian of the reaction part evaluated at the fixed point $(u^*,v^*)$. The generalised Laplacian operator  $\mathcal{L}$ is given by
\begin{equation}
\mathcal{L}=\left(\begin{matrix}D_u \mathbf{L} & \mathbf{0}_N\\ \mathbf{0}_N &D_v \mathbf{L}\end{matrix}\right)=\left(\begin{matrix}D_u  & 0 \\ 0 &D_v \end{matrix}\right)\otimes \mathbf{L}\, ,
\label{eq:L}
\end{equation}

The eigenvalues $\lambda$ of the $2N \times 2N$ matrix $\mathcal{J}_0+\mathcal{L}$, define the fate of the perturbation. If there exists (at least) one eigenvalue with a positive real part, then the perturbation initially grows exponentially, and the homogeneous fixed point is predicted unstable, as follows the conventional Turing mechanism. An elegant procedure to compute the spectrum of $\mathcal{J}_0+\mathcal{L}$, involves the eigenvectors of the Laplacian operator $\mathbf{L}$. Assume the latter to be diagonalisable, introduce its  eigenvalues $\Lambda^{(\alpha)}$ and associated eigenvectors $\phi^{(\alpha)}$, as $\sum_{j} L_{ij} \phi_j^{(\alpha)} = \Lambda^{(\alpha)} \phi_i^{(\alpha)}$, with $\alpha=1,...,N$. Then we can expand the perturbation on the basis of the Laplacian operator and project the $2N \times 2N$ Eqs.~\eqref{eq:RDLin}, onto a collection of $N$ independent $2 \times 2$ problems, each referred to a given subspace, as spanned by the corresponding eigenvector. When the underlying network is symmetric, the Laplacian eigenvalues $\Lambda^{(\alpha)}$ are real and non-positive. Moreover, the eigenvectors form an orthonormal basis of the embedding manifold. The short time evolution of the imposed perturbation is exponential and the associated  growth (or damping) factors $\lambda$ can be readily computed as a function of the  eigenvalues entries $\Lambda^{(\alpha)}$, by solving the following characteristic problem:
\begin{equation}
\label{det}
\det \left(\begin{matrix}f_u+D_u \Lambda^{(\alpha)} - \lambda & f_v \\ g_u &g_v +D_v \Lambda^{(\alpha)} - \lambda  \end{matrix}\right)=0
\end{equation}
The eigenvalue with the largest real part,  $\lambda =\max_{\alpha} \Re \lambda (\Lambda^{(\alpha)})$, defines the so called dispersion relation, which characterises the response of the deterministic system (\ref{eq:RD}) 
to external perturbations. The quantity $\Lambda^{(\alpha)}$ is the analogue of the wavelength for a spatial pattern in a system defined on a continuous regular lattice, where $\Lambda^{(\alpha)} \equiv - k^2$, and $k$ stands for the usual spatial Fourier mode. When the dispersion relation $\lambda (\Lambda^{(\alpha)})$ is positive over a finite range of $\Lambda^{(\alpha)}$, the perturbation is triggered unstable. The combined action of the reactive and diffusive components produces a self-consistent amplification of the small inhomogeneities, driving the system towards a patterned stationary stable attractor. This constitutes the natural extension of the celebrated Turing patterns, to systems  defined on a discrete, network-like support. The discrete dispersion relation is indeed identical to the one for continuous support, except for the fact that its domain of definition takes values on the finite set of real eigenvalues $\Lambda^{(\alpha)}$. The conditions for the onset of the instability in the continuum and discrete settings hence coincide, for  sufficiently large networks and modulo small deviations that might eventually arise, if the discrete spectrum does not extend to the region where the dispersion relation for the continuous support is positive. 

A completely different scenario is, instead, faced when networks are assumed to be directed, e.g. when the edges connecting two adjacent nodes can be accessed in one direction only.  Breaking the symmetry, implies dealing with complex  eigenvalues of the associated Laplacian, a modification that can significantly impact the onset of diffusion-driven, Turing-like, instabilities. Indeed, the imaginary component of $\Lambda^{(\alpha)}$ contributes to the real part of $\lambda$, as follows the self-consistent condition (\ref{det}). It can be shown that this additional contribution favours the outbreak of the instability,  thus enhancing the intrinsic ability of the reaction-diffusion system to develop ordered patterns, as compared to its continuum (or discrete and symmetric) counterpart~\cite{asllani}. 

In this paper we are interested in the emergence of patterns in systems defined on an asymmetric support, which are deemed stable under a conventional linear stability analysis, for any kind of perturbations. The latter can in fact grow, at short times, if the underlying network is non-normal, yielding patterns also when linear stability returns a negative dispersion relation, $\lambda<0$, a setting where Turing instability cannot develop. The remaining part of this section is devoted to revisiting the concept of non-normality. We will in particular discuss how non-normality impacts the evolution of a generic linear system, in arbitrary dimensions. To this end consider a linear system $\frac{d}{dt}\mathbf{x}= \mathbf{M} \mathbf{x}$, and assume $\mathbf{M}$ to be non-normal. A matrix is said to be non-normal, if it does not commute with its adjoint~\cite{trefethen}. If $\mathbf{M}$ is real, taking the adjoint amounts to computing the transpose of the matrix. Hence, $\mathbf{M}$ is non-normal provided $[\mathbf{M},\mathbf{M}^T] \equiv \mathbf{M} \mathbf{M}^T - \mathbf{M}^T \mathbf{M} \neq 0$, where the apex $T$ stands for the transpose operation. Observe that non-normality is equivalent to the non-existence of a suitable orthogonal basis of eigenvectors.

A straightforward manipulation \cite{trefethen} yields the following equation for  the evolution of the norm of the perturbation 
$\Vert{\delta\mathbf{x}}\Vert$:
\begin{equation}
\frac{d \Vert{\delta \mathbf{x}}\Vert}{d t}=\frac{\delta\mathbf{x}^T\mathbf{H}{(\mathbf{M})}\delta\mathbf{x}}{\Vert{\delta \mathbf{x}}\Vert}
\end{equation}
where $\mathbf{H}{(\mathbf{M})}=\frac{\mathbf{M}+\mathbf{M}^T}{2}$ identifies the Hermitian part of $\mathbf{M}$. The evolution of the perturbation at short times is ultimately set by 
the so called {\it numerical abscissa}, $\omega=\sup\sigma\bigl(\mathbf{H}{(\mathbf{M})}\bigr)$, where $\sigma\bigl(\mathbf{H}(\mathbf{M})\bigr)$ stands for the spectrum of $\mathbf{H}{(\mathbf{M})}$. If $\omega>0$, then the non-normal matrix is termed {\it reactive}, and perturbations may display an initial, transient growth. The degree of non-normal amplification can be quantified through diverse indicators \cite{trefethen}. In our setting, the reaction diffusion system is amenable to its linear homologue (\ref{eq:RDLin}), when operating in the vicinity of a stationary stable homogeneous attractor. Asymmetry, a key ingredient for non-normality, can be hence accommodated in the matrix that encodes for paired couplings, so triggering a short time amplification of the imposed perturbation that will prove instrumental for a generalised class of diffusion-driven instabilities.

\section{Non-normal patterns for stable systems} 

\begin{figure}[!ht]
	\centering
		\includegraphics[scale=0.38]{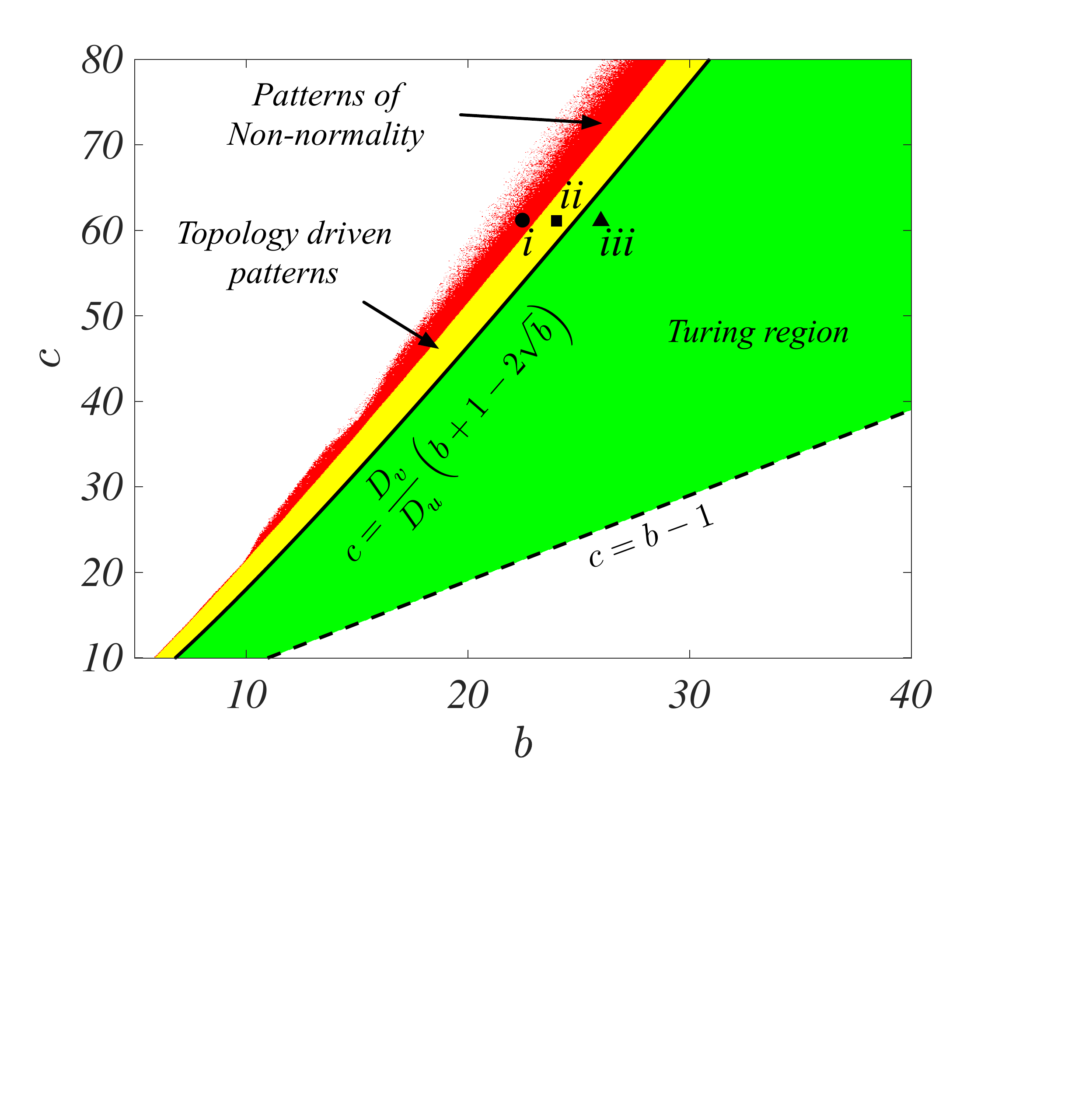}
		\vspace{-3.5cm}
	\caption{\textbf{Different domains in the parameter plane $(b,c)$ that yield pattern formation for the Brusselator model.} Distinct domains of interest are isolated in the reference plane $(b,c)$ and depicted with different colour codes. In region $(iii)$ (shaded in green), a conventional Turing instability develops for the Brusselator model defined on a symmetric support. In region $(ii)$ (yellow), topology driven patterns emerge: the homogeneous fixed point is triggered unstable by the directed nature of the spatial support; here patterns can be stationary or wave-like. Finally, red dots define region $(i)$, where non-normality drives the onset of pattern formation, notwithstanding the prediction of the linear stability analysis that deems the homogeneous fixed point stable, and thus resilient to  external perturbation. In white is the region where patterns cannot be established. The Turing region is bounded from below by the curve $c=b-1$ (dashed line) and from above by $c=\frac{D_v}{D_u}(b+1-2\sqrt{b})$ (solid curve).  The underlying non-normal network (generated according to the procedure discussed in the main body of the paper) is made up of $N=100$ nodes. The model parameters are $D_u = 0.5$ and $D_v = 1.925$. the initial conditions for $u$ (resp. $v$) are uniformly drawn from an $N$-dimensional sphere of radius $\delta = 0.2$ centred $u^*$ (resp. $v^*$).}
	\label{fig:NNPatt}
\end{figure}

Without loss of generality, we shall assume in the following the Brusselator model~\cite{PrigogineNicolis1967,PrigogineLefever1968,McKane2008}, as a reference reaction scheme. The Brusselator is  
a paradigm of non-linear dynamics, and it is often employed in the literature as a reference model for self-organisation, synchronisation and pattern formation. Our choice amounts to setting 
$f(x,y)=1-(b+1)x+cx^2y$ and $g(x,y)=bx-cx^2y$ where $b$ and $c$ denote positive parameters. The system admits a trivial homogeneous fixed point for $(u_i,v_i)=(1,b/c)$, $\forall i$, that is stable provided $c>b-1$. We can easily determine the conditions for a Turing instability. Assuming as a starting point a symmetric spatial support, we conclude that the diffusion coefficients need to comply with the necessary condition $D_u<D_v$, for patterns to emerge. A straightforward calculation enables us to conclude that $c<D_v/D_u (b+1-2\sqrt{b})$, as an additional constraint on the parameters, for the Turing instability. The above conditions allow us to isolate the domain in the parameter space $(b,c)$ where Turing patterns are predicted to occur. The region of interest is displayed in Fig.~\ref{fig:NNPatt}, with a green shading, and it is delimited by the curves $c=b-1$ (dashed line) and $c=D_v/D_u (b+1-2\sqrt{b})$ (solid line) (region $(iii)$).

Consider now the Brusselator model defined on a directed, non-normal network. In the following we choose to operate with the class of non-normal networks introduced in~\cite{nonnormal}. The recipe for the network 
generation goes as follows. Start with a weighted directed $1D$ ring, and assume the weights to be randomly drawn from a uniform distribution $U[0,\gamma]$, where $\gamma$ is a (large) scalar parameter. 
Further, label the nodes with an ascending index, when circulating clockwise across the ring. Then, for all $i = 1,\dots ,N$, we create with probability $p_1$, $0 < p_1 < 1$, {a weak link of order $1$, namely much smaller than the weights on the directed ring,} that bridges the $(i + 1)$-th node to the $i$-th one. This procedure returns a non-normal network which displays a richer topology as compared to that of the initial direct skeleton. Moreover, its degree of non-normality correlates positively with $\gamma$: the larger $\gamma$ the more pronounced the inherent non-normality, as revealed through standard quantitative indicators~\cite{nonnormal}. In the following we shall operate with $\gamma=50$ and $p_1=0.4$.

As already stated, the imaginary component of the eigenvalues of the Laplacian operator {seeds an enlargement} of the region of parameter space where the instability is predicted to occur. The portion of the parameter plane where the instability extends is depicted in yellow in Fig.~\ref{fig:NNPatt} and labeled as region $(ii)$. Here, the dispersion relation is positive and the ensuing patterns bear a topological imprint, as they are ultimately reflecting the directionality of the embedding spatial support. The boundaries of the region where topological patterns develop can be traced analytically~\cite{asllani}.

More interesting for our current study, is what happens above region $(ii)$. Here, the dispersion relation, the rate which governs the exponential evolution of the perturbation, is negative, thus implying linear stability. Patterns, resembling those generated inside the domain of (topological or Turing-like) instability, are instead observed, when integrating numerically the governing equations. A red dot is plotted in Fig.~\ref{fig:NNPatt} if the patterns are obtained, upon numerical integration, for the corresponding values of the reaction parameters $(b,c)$. The results displayed in Fig.~\ref{fig:NNPatt} refer to one realisation of the dynamics. Repeating the analysis to account for an extended set of independent numerical experiments yield equivalent conclusions.  The amplitude of the noise is assumed so small that patterns cannot set in when the non-normal network is replaced with its symmetrised analogue. Denote by $\mathbf{A}$ the adjacency matrix of the non-normal network, as obtained via the procedure discussed above. Then, $(\mathbf{A}+\mathbf{A}^T)/2$ is the adjacency matrix which characterises the associated symmetric network. We anticipate that 
propensity to self-organisation exhibited by the system, stems ultimately from the {characteristic} dynamics as displayed by linear non-normal systems at short times. We shall return to elaborate 
on this subject in the following. 

\begin{figure*}[!ht]
	\centering
	\includegraphics[scale=0.25]{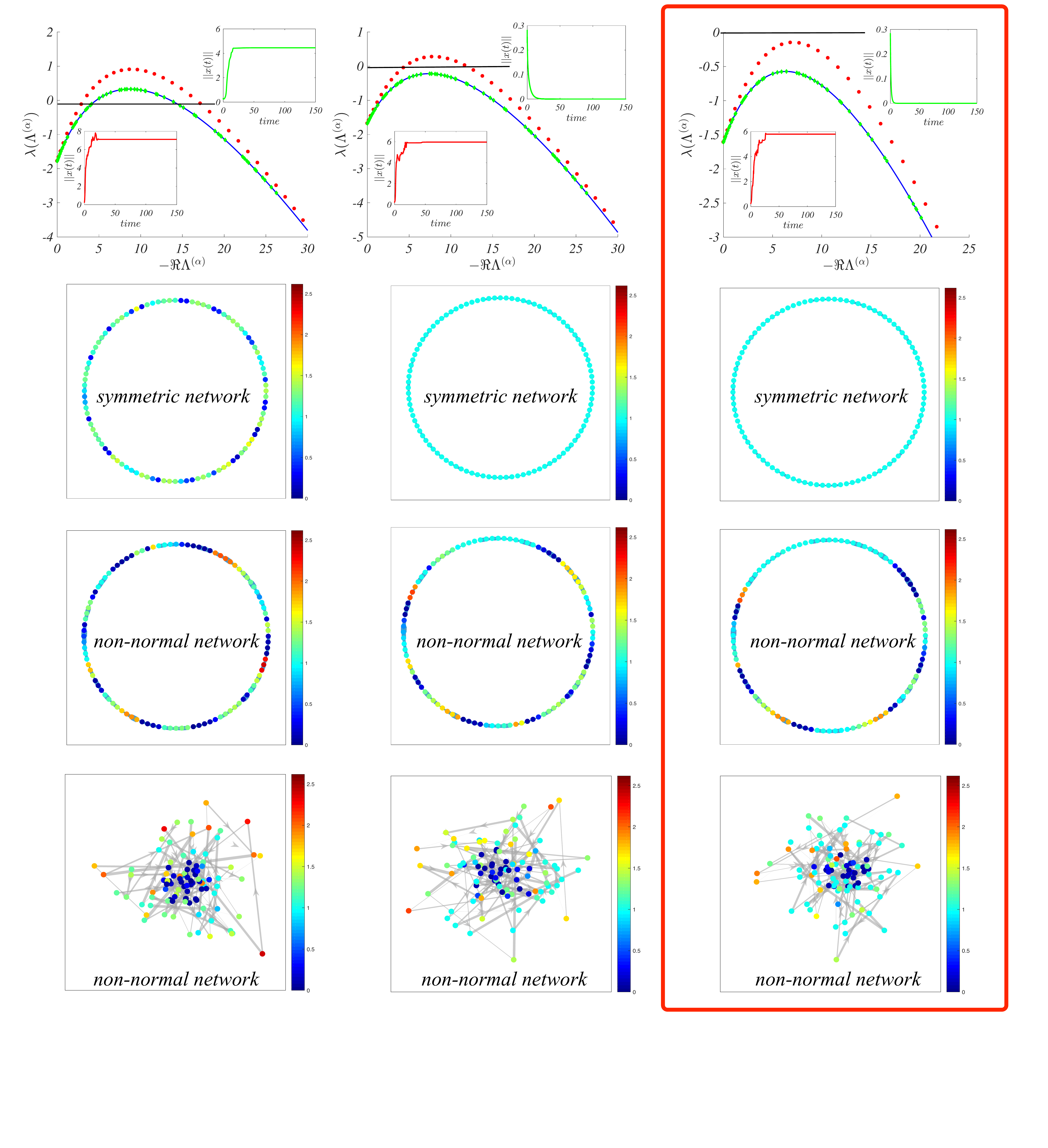}
	\vspace{-1cm}
	\caption{\textbf{The dispersion relation and the ensuing patterns in different regions of the parameter plane.} Turing patterns $(b,c)=(26,61.1)$ (left column), topology-driven patterns $(b,c)=(24,61.1)$ (middle column), patterns of non-normality $(b,c)=(22.5,61.1)$ (right column, enclosed in the red box). The position of the selected working points is marked in Fig.~\ref{fig:NNPatt}, with, respectively, a triangle, a square and a circle. In all cases, the dispersion relation is drawn for the non-normal (directed) network (red dots) and the symmetrised network (green dots). Recall that the instability is set once the  dispersion relation is positive. The insets in the panels that form the first row, show the time evolution of the norm of the system for the non-normal network (red curves) and the symmetric settings (green curves). Here, the system is initialised $\delta$-close to the equilibrium (see caption of Fig.~\ref{fig:NNPatt}), with $\delta=0.2$. The remaining parameters, as well as the network employed, are those used in Fig.~\ref{fig:NNPatt}. {The ensuing patterns are displayed for, respectively, the symmetric setting (second row), and the non-normal network. In particular, in the third row, the patterns are depicted when representing the network with a stylized lattice layout (as for the symmetric setting). In the fourth row, the edges of the networks are instead shown. Nodes displaying a low concentration are assigned to the central portion of the cluster. }}
	\label{fig:RelDispPatt}
\end{figure*}

In Fig.~\ref{fig:RelDispPatt} the dispersion relation is shown for different choices of the parameters $(b,c)$.  Selected working points are displayed with a symbolic marker (respectively circle, square, triangle) in Fig.~\ref{fig:NNPatt}.
In the top-left panel of  Fig.~\ref{fig:RelDispPatt} the dispersion relation $\lambda$ is plotted against the real part of the Laplacian eigenvalues $\Lambda$, when operating inside the region of conventional Turing order. The continuum dispersion relation (solid blue line) is hence unstable, i.e. it assumes positive values. Green symbols, obtained for the Brusselator model evolved on a symmetrised network support (as defined above), follow, as expected, the continuum profile. Red circles refer instead to the model operated on a non-normal network, of the type discussed above. Again the system is predicted unstable, the estimated  growth rate of the perturbation being larger as compared to the symmetric case. In the insets annexed to the panel, the time evolution of the norm of the perturbation is represented, for both symmetric (top, green curve) and asymmetric (down, red curve) settings. In both cases the curves grow and eventually reach an equilibrium plateau, the signature of the existence of a non-homogeneous solution. The corresponding patterns, 
as obtained in the asymptotic regime of the evolution, are displayed in the boxes annexed right below the corresponding dispersion relation. The nodes of the collection, arranged in a circular array, are coloured with an {appropriate} code which reflects the steady state concentration of species $u$. The system segregates into nodes rich/poor of activators/inhibitors, the polarisation being more notable in the asymmetric setting (bottom figure). In the second plot of the top row, the parameters are set so as to have the system initiated in region $(ii)$ of Fig.~\ref{fig:NNPatt} (square symbol). The system defined on symmetric graphs is now stable: the norm of the perturbation damps steadily and the homogeneous fixed point proves resilient to exogenous perturbation of the initial state (the dispersion relation is negative). No patterns can hence develop, as displayed in the second panel of the second row. Conversely, the dispersion relation computed for the Brusselator on the directed network (red circles) signals the instability, which eventually materialises in a topology driven pattern, as displayed in the bottom middle panel. Finally, for region $(i)$ we obtain the result enclosed in the red box of Fig.~\ref{fig:RelDispPatt}. The dispersion relations are now negative, thus implying that the perturbation should asymptotically vanish. While no patterns are observed for the system integrated on a symmetric support, this is manifestly not the case when the underlying graph is assumed to be non-normal. The short time growth of the perturbation, as reported in the small (lower) inset of the top right panel, stabilises in a macroscopic pattern (lower panel of the column enclosed in red) which shares striking similarities (in terms of density distribution and corresponding amplitude) with the homologous patterns obtained inside the region of linear instability. From these observations it appears already evident that the mechanism that drives the emergence of non-normal pattern from the asymptotically stable homogeneous solution resides in the ability to amplify the perturbation imposed at short time. The idea, that we will now explore, is that the short time amplification, as triggered by non-normality, makes the system jump across the barrier that separates homogeneous and heterogeneous attractors, thus yielding a spatially extended pattern, which could not be anticipated via a linear stability analysis.  

\begin{figure*}[t]\vspace*{-.5cm}
	\centering
	\includegraphics[scale=0.35]{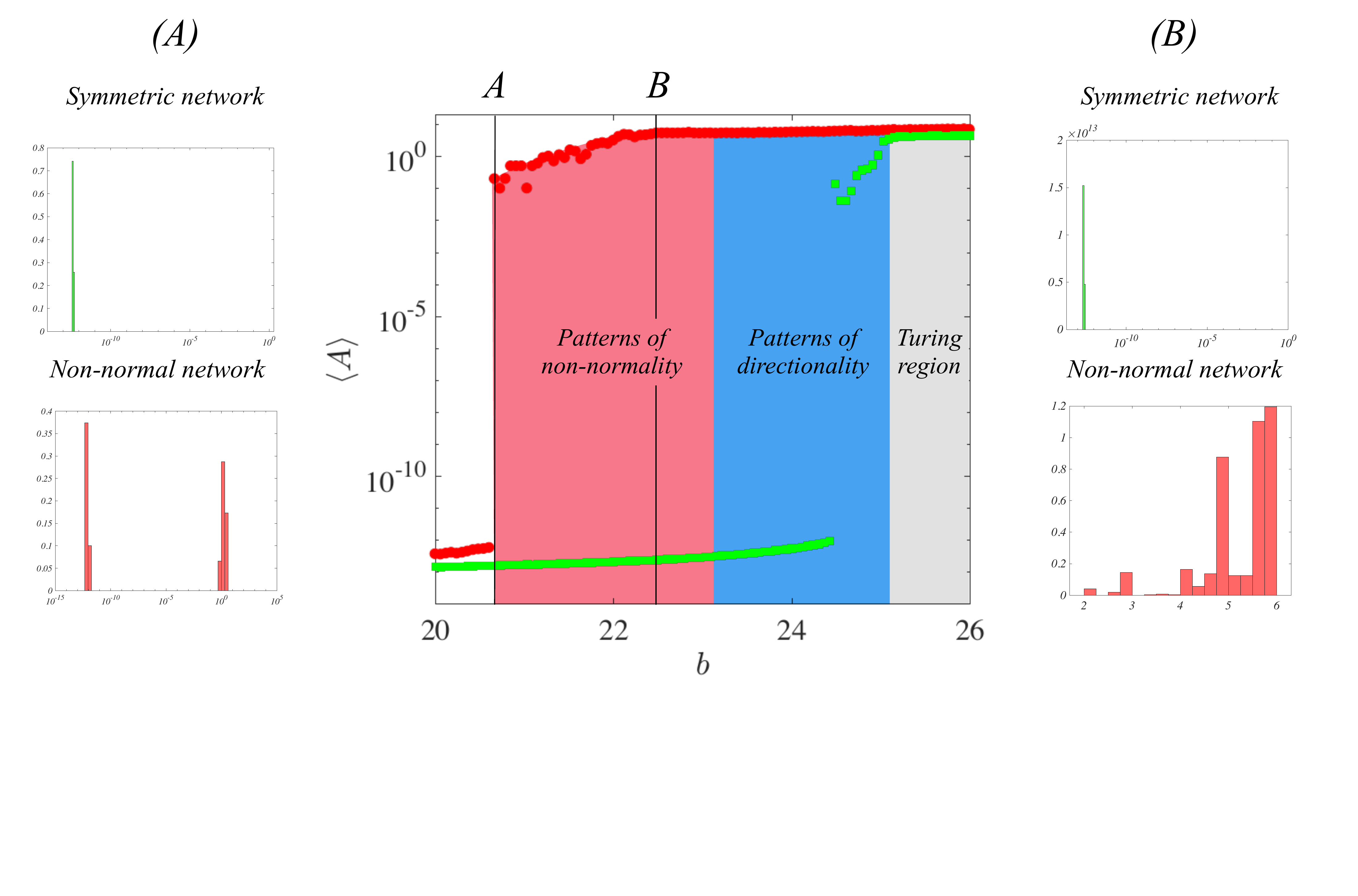}\vspace*{-2.5cm}
	\caption{\textbf{On the amplitude of the ensuing patterns.} Main panel (centre): the average pattern amplitude $\langle A \rangle$ is plotted as a function of the parameter $b$, for $c=61.1$. Red circles refer to the system evolved on a non-normal network, while green squares stand for its symmetrised analogue. The grey region corresponds to parameters giving rise to Turing instability and the blue one to instability driven by the network directionality. Finally, the domain coloured in pink highlights the region where the homogeneous fixed point is stable according to a linear stability analysis, but where patterns can emerge due to non-normality. The insets show histograms of pattern amplitudes, for two specific values of $b$,  namely $b\sim 20.5$ (A), $b\sim 22.5$ (B). Red symbols refer to data collected when the Brusselator model is evolves on a non-normal support. Green histograms are computed from simulations that assume symmetric support. The remaining parameters have been set to $D_u=0.5$, $D_v=1.925$, $\delta=0.2$. Simulations have been performed using the same networks as in Fig.~\ref{fig:NNPatt}.}
\label{fig:NNPattsections}
\end{figure*}

To further investigate this issue, we introduce a quantitative indicator, $A$, for measuring the amplitude of the patterns that are eventually established, by means of the alternative routes highlighted above. This is defined as:
\begin{equation}
A=\sqrt{\sum_i\left[(u_i^{\infty}-u^*)^2+[(v_i^{\infty}-v^*)^2\right]}\, ,
\label{eq:pattamp}
\end{equation}
where $u_i^{\infty}$ (resp. $v_i^{\infty}$) is the asymptotic stationary value of $u_i(t)$ (resp. $v_i(t)$) at node $i$. In Fig.~\ref{fig:NNPattsections}, the scalar quantity $A$ averaged over a large sample of independent realisations, $\langle A \rangle$, is plotted as a function of $b$, at fixed value of $c$. Circles (red) refer to the system evolved on the non-normal network, the same network employed in the analysis that led to Fig.~\ref{fig:NNPatt}. Squares (green) stand for the values of $\langle A \rangle$ recorded when the system is studied on the symmetrised version of the network. As can be appreciated by visual inspection of Fig.~\ref{fig:NNPattsections}, non-normality drives  a significant enlargement of the region of the parameter space that yields heterogeneous patterns, beyond the domain where topological patterns (or pattern of directionality) are found to occur. Notice also that the patterns sustained by non-normality have a characteristic amplitude comparable to that associated with conventional Turing patterns, as quantified by $\langle A \rangle$. The same conclusion is reached upon inspection of the annexed histograms. Performing a vertical cut, for a value of $b$ at the threshold (vertical line located at the point A) of the leftmost bifurcation, one finds that non-normal patterns are characterised by a bimodal distribution of the associated amplitude. The coexistence of homogeneous, $\langle A \rangle\sim 0$, and patterned solutions, $\langle A \rangle=\mathcal{O}(1)$, suggests that the transition is of first order. The histogram obtained for a symmetric setting returns instead a unique peak, at zero $\langle A \rangle$. This is also true when the vertical cut is performed well inside the region of non-normal patterns (vertical line located at B). Non-normal patterns are instead associated with macroscopic amplitudes.  

The results of Fig.~\ref{fig:NNPattsections} reveal another important fact: patterns are also found  for symmetric networks, in a small region outside (but in proximity of) the Turing domain. Such patterns are again due to non-normality, as stemming from the reaction term. Indeed the Jacobian of the Brusselator is non-normal for any choice of $b$ and $c$, and this is a sufficient condition to trigger the system unstable, for a perturbation of modest size, also beyond the strict boundaries, obtained under the linear stability framework. The positive interference of two distinct sources of non-normality, respectively associated with the reaction and diffusion parts, brings {an enlargement} of the region deputed to the instability, as already remarked. Interestingly enough, patterns of non-normality can also exist for systems displaying a symmetric Jacobian, when the embedding spatial support is made non-normal ~\cite{nonnormal}.  

\begin{figure*}[t]\vspace*{-2cm}
	\centering
	\includegraphics[scale=0.35]{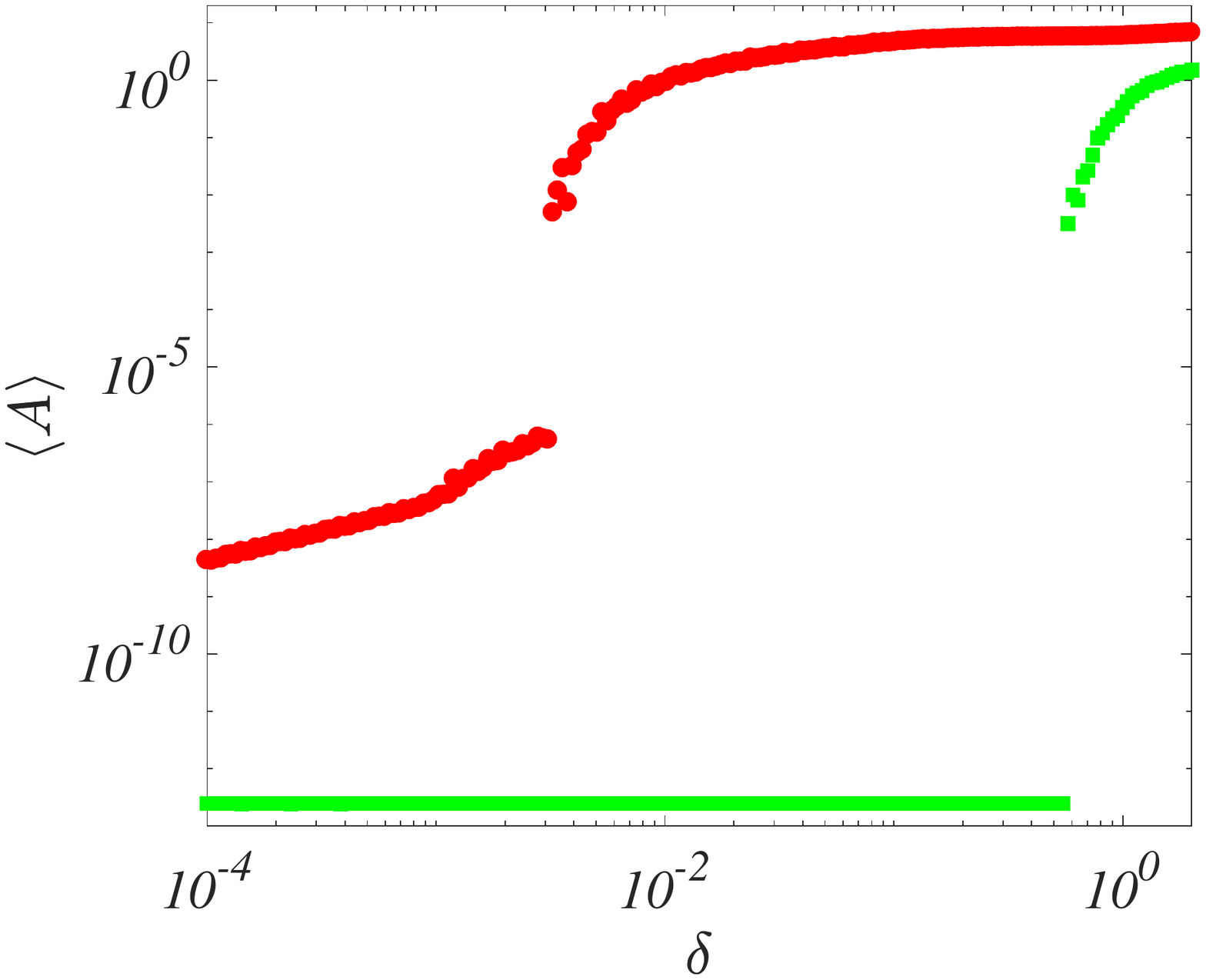}\quad
	\includegraphics[scale=0.35]{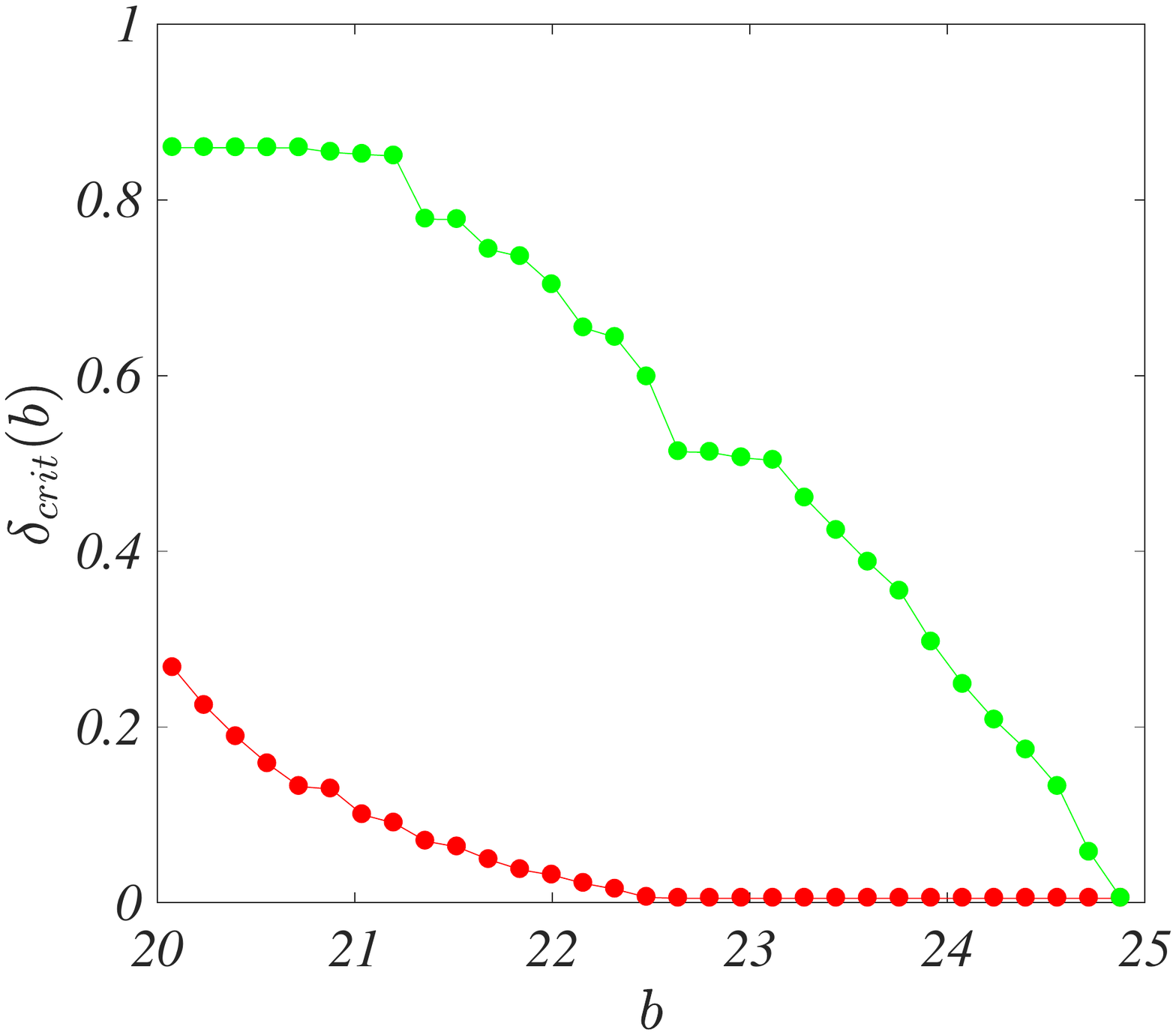}
	\vspace{-2cm}
	\caption{\textbf{On the stability domain.} Left panel: average pattern amplitude as a function of the size of the initial perturbation $\delta$, for $b=22.56$. The average is computed over $500$ independent realisations of the dynamics. Right panel:  the size of the stability domain $\delta_{crit}$ is computed for, respectively, the Brusselator defined on the non-normal network (red points) and on its symmetric analogue (green points), and plotted as a function of $b$. The remaining parameters are set to $c=61.1$, $D_u=0.5$, $D_v=1.925$. For the analysis we employed the networks used in Fig.~\ref{fig:NNPatt}.}
	\label{fig:NNPattDelta}
\end{figure*}

\begin{figure*}[t]\vspace*{-2cm}
	\centering
	\includegraphics[scale=0.35]{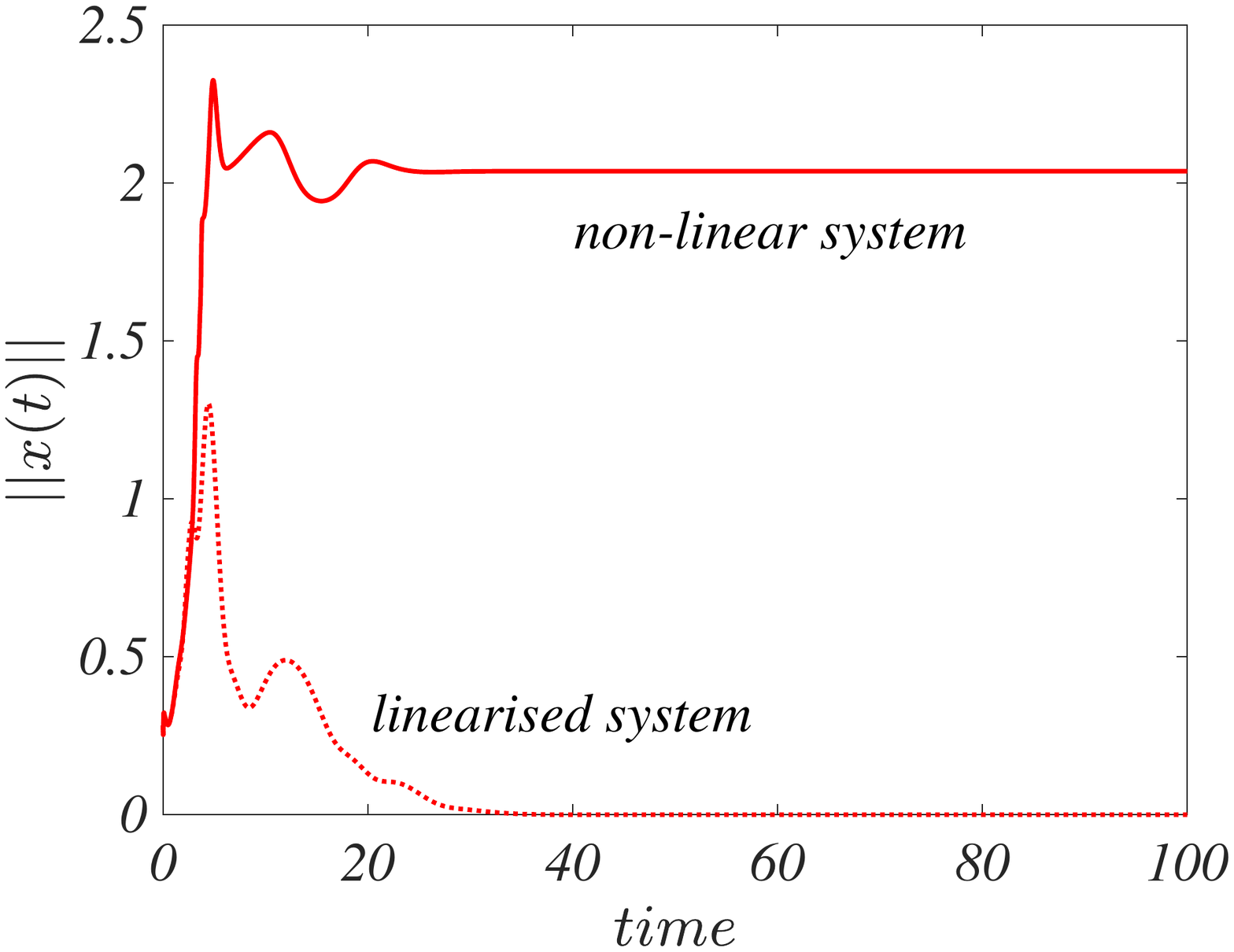}\quad
	\includegraphics[scale=0.35]{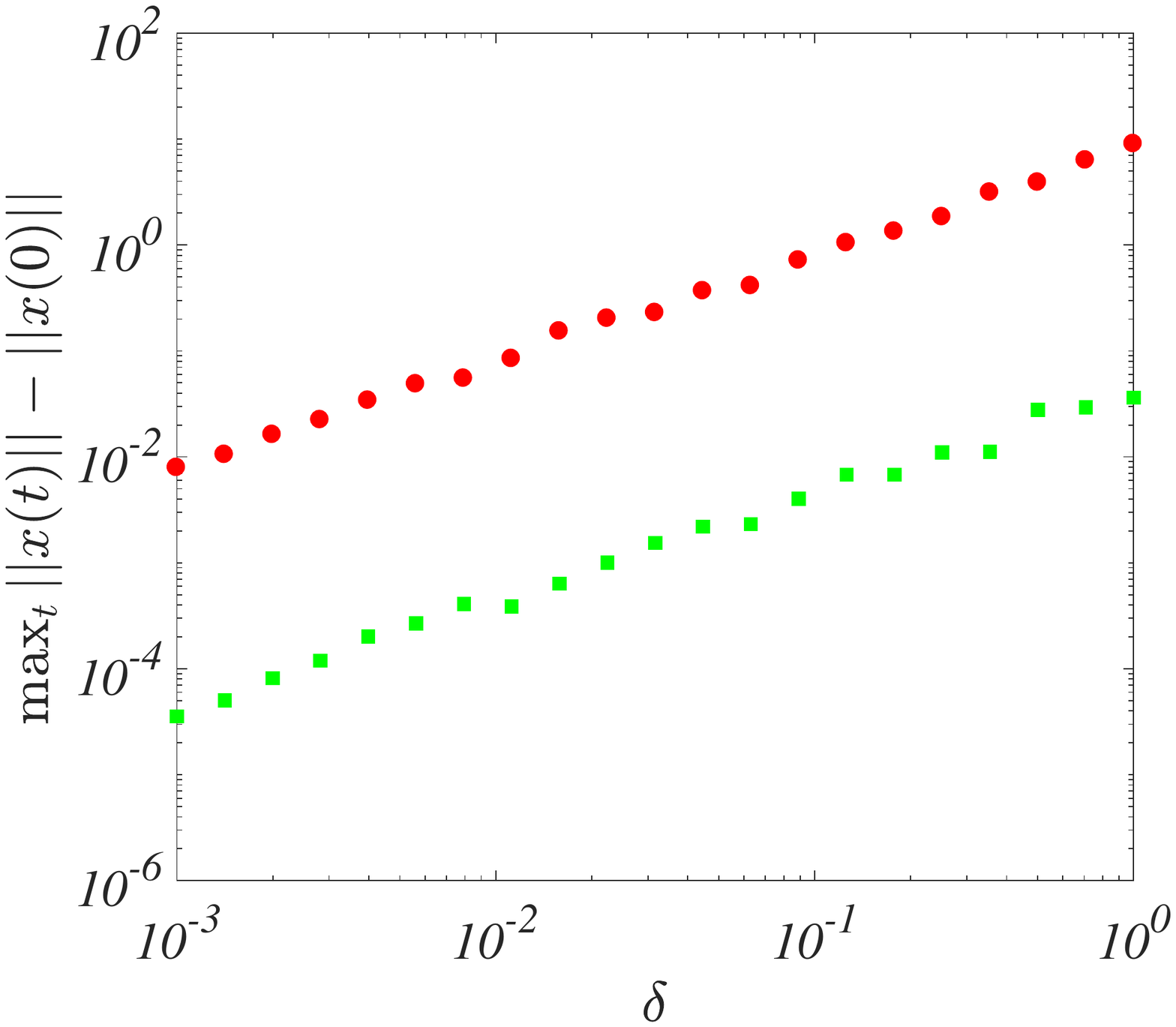}
	\vspace{-2cm}
	\caption{\textbf{Transient amplification and linearised dynamics.} Left panel: we show the time evolution of the {norm} of the perturbation for the linearised dynamics (dotted line) and for the full non-linear model (solid line). For the linear system, we can see the transient growth and the subsequent decrease towards the asymptotic equilibrium. The non-linear system stabilises eventually to a value of the norm which is different from zero. Here, $\delta=0.2$. Right panel: $\Delta(\delta):=\max_t ||\vec{x}(t)||-||\vec{x}(0)||$ as a function of the perturbation size $\delta$, for non-normal (red dots) and symmetric networks (green dots). The  parameters are set to $b=22$, $c=61.1$, $D_u=0.5$, $D_v=1.925$. The networks employed are those used in Fig.~\ref{fig:NNPatt}.}
	\label{fig:LinDyn}
\end{figure*}

To shed further light on the mechanism that seeds the instability at negative dispersion relation, and so elaborate on the role played by non-normality, we study the response of the system, on both symmetric and asymmetric supports, at different values of the size of the initial perturbations applied on the homogeneous equilibrium. In the left panel of Fig.~\ref{fig:NNPattDelta}, we plot the average patterns amplitude as a function of  $\delta$, for a given choice of $b$ : in the case of symmetric networks (green points) patterns can only develop for macroscopic perturbations ($\delta > \delta_{crit} = 0.83$).  On the other hand, for non-normal networks, a perturbation with $\delta > \delta_{crit} =0.005$ suffices for the dawning of the patterns. For any $b$ we define $\delta_{crit}$ to be the smallest value of the initial perturbation for which patterns emerge. In the right panel of  Fig.~\ref{fig:NNPattDelta}, the value of $\delta_{crit}$ is represented against $b$: the basin of attraction of the homogeneous state shrinks  considerably, when the system is defined on a non-normal support, an observation which contributes to explain the augmented propensity of non-normal systems for pattern formation. 
The shrinking of the basin of attraction of the homogeneous fixed point {follows} the transient growth of the perturbation, as driven by non-normality. In the left panel of Fig~\ref{fig:LinDyn} the evolution of the norm of the perturbation is tracked under the linear approximation, which follows Eq. (\ref{eq:RDLin}). The norm grows at short times, and then converges to zero, as it should since the homogeneous fixed point proves linearly stable. The evolution of the perturbation as obtained for the full non-linear model is instead quite different, as can be visually appreciated. Non-linearities eventually stabilise the norm to a non-zero value which was made accessible to the system under the transient evolution. In the right panel of Fig. \ref{fig:LinDyn}, $\Delta(\delta):=\max_t ||\mathbf{x}(t)||-||\mathbf{x}(0)||$ is represented as a function of the perturbation size $\delta$, for respectively, non-normal network (red symbols) and symmetric one (green symbols). The two curves run parallel to each other in log-log scale and the relative shift is rationalised as follows: Consider, for example, the right panel of Fig. \ref{fig:NNPattDelta}, where $\delta_{crit}$ is plotted against $b$ and focus in particular on $b=22$. For the case of a non-normal network, we readily obtain
 $(\delta_{crit})_{nn} \simeq 0.05$, while $(\delta_{crit})_{sym}  \simeq 0.6$. On the other hand, the value of $\Delta(0.05)$ (i.e. a direct measure of the transient growth induced by non-normality for $\delta \simeq 0.05$, under the linear approximation) is about $0.4$ (see right panel of Fig. \ref{fig:LinDyn}, drawn for $b=22$), i.e. of the correct oder of magnitude to explain the observed discrepancy between $(\delta_{crit})_{sym}$ and $(\delta_{crit})_{nn}$.

\begin{figure}[t]
	\centering
	\includegraphics[scale=0.35]{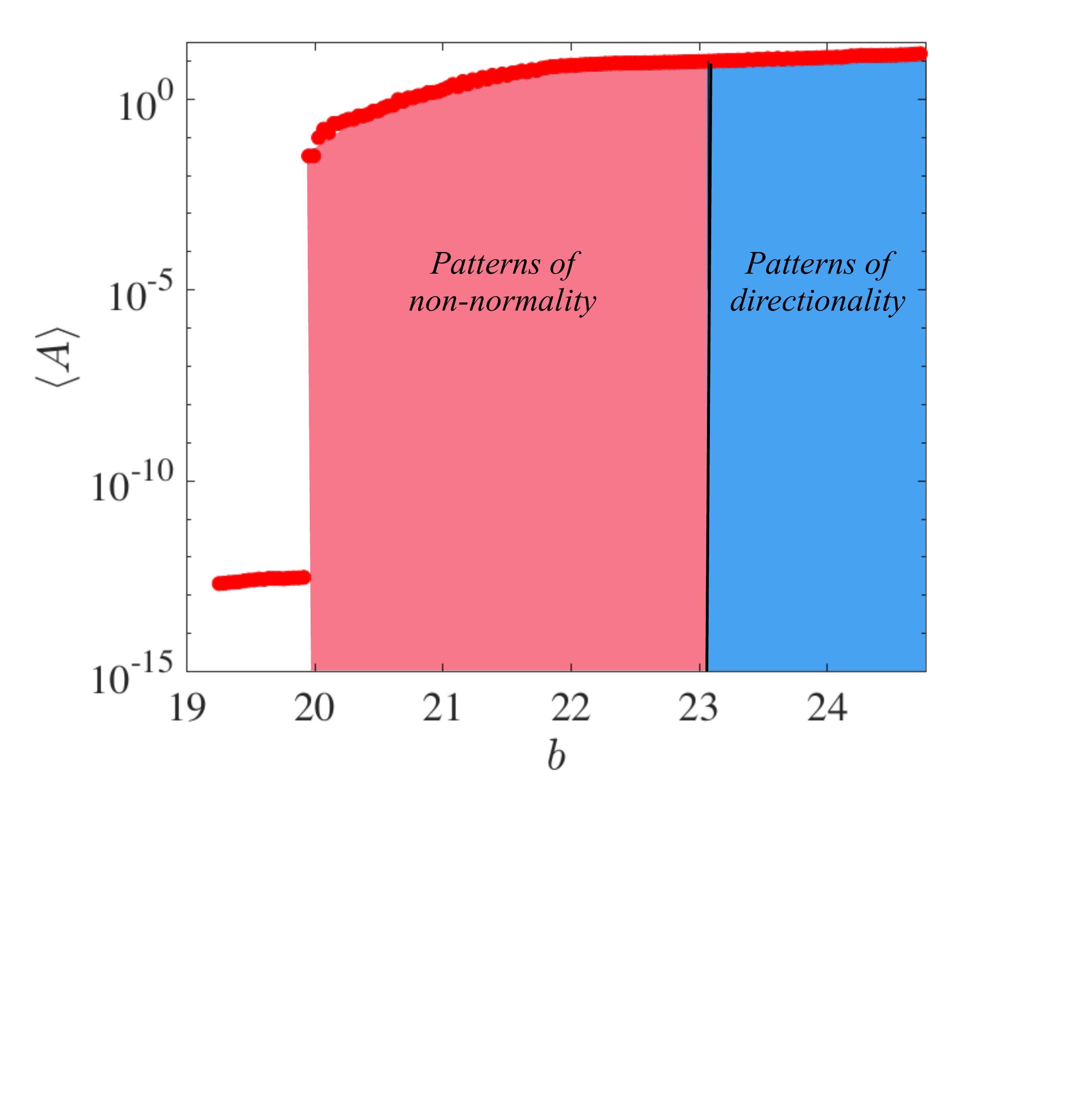}
	\vspace{-2cm}
	\caption{\textbf{Patterns of non-normality for $D_v/D_u\sim 1$.} 
We report the average pattern amplitude $\langle A \rangle$ as a function of the parameter $b$, for $c = 23.72$ and $D_v/D_u=1.5$, for the Brusselator model evolving on the non-normal network used in Fig.~\ref{fig:NNPatt}. Values of $b$ lying in the blue region yield instability driven patterns due to the network directionality. The region coloured in pink highlights the interval in $b$ where the homogeneous fixed point is linearly stable and patterns can emerge due to non-normality. Each point is the average over $100$ independent replicas. The remaining model parameters are the same as those used in Fig.~\ref{fig:NNPatt}.}
	\label{fig:DvDusim1}
\end{figure}

As recalled in the Introduction, Turing patterns requires the inhibitor to diffuse faster than the activator. {Moreover, the ratio of the diffusivities $D_v/D_u$ should be sufficiently large} for the pattern to materialise in a extended region of parameters. For the Brusselator,  sending $D_v/D_u \rightarrow 1$ contextually implies setting $b$ and $c$ very large for the patterns to set in, via a conventional Turing instability. Network asymmetry, and the associated non-normality, enable us to extend the region where patterns are reported to occur to smaller ratios of $D_v/D_u \rightarrow 1$, without forcing the reactive parameters to unphysical large values. As an example see Fig.~\ref{fig:DvDusim1} where we show the pattern amplitude for $D_v=0.75$ and $D_u=0.5$, resulting in $D_v/D_u=1.5$, for values of the parameters $b$ and $c$ of the same order of the ones used in Fig.~\ref{fig:NNPattsections}. Patterns of non-normality do exist for a large range of values of the $b$ parameter (observe also that in this setting Turing patterns cannot develop on a symmetric support). The patterns are of amplitude with a magnitude up to $b\sim 21$. Below this value the magnitude reduces, as depicted in Fig.~\ref{fig:DvDusim1}. This is a spurious effect due to the fact that, for $b \in [20,21]$, there is a coexistence of two distinct solutions, one corresponding to the macroscopic pattern and the other to the persisting homogeneous state (see also Fig.~\ref{fig:NNPattsections}).

\section{pseudo-dispersion relation}
\label{sec:meaning}
In the above discussion we have shown that patterns can emerge due to the non-normality of the underlying network, when the dispersion relation would deem the homogeneous fixed point stable. Our conclusions rely on a direct numerical integration of the governing equations and we lack a predictive tool that could eventually be employed to anticipate the phenomenon. In the following we shall propose an extended definition of the dispersion relation that builds on the concept of the pseudospectrum~\cite{trefethen}. We begin by recalling the definition of the pseudospectrum $\sigma_\epsilon(\textbf{M})$  of, a matrix $\textbf{M}$, namely:
\begin{equation}
\sigma_\epsilon(\textbf{M})=\sigma(\textbf{M}+\textbf{E}),\; \forall\;||\textbf{E}||<\epsilon\, ,
\label{eq:pseudo}
\end{equation}
where $\sigma(\textbf{M})$ denotes the spectrum of the matrix $\textbf{M}$ {and $\epsilon$ is a scalar positive defined quantity}.
The pseudospectrum is an important and relatively recent mathematical tool which has been successfully applied to different disciplinary contexts where non-normality {occurs} \cite{trefethen}. The Kreiss constant, a lower bound for the linear growth of an imposed perturbation, can be computed from the pseudospectrum.

While the  behaviour of a linear (non-normal) system is relatively well understood in the framework of pseudospectrum theory,  extending this to non-linear models is challenging. 
Consider a generic non-linear system, $\dot{\textbf{x}} = \textbf{f}(\textbf{x})$, where $\textbf{f}: \mathbb{R}^n \rightarrow \mathbb{R}^n$ is a vector-valued non-linear function. 

The problem of local stability is tackled by performing a Taylor expansion about a given fixed point $\textbf{x}^*$, $\textbf{f}(\textbf{x}^*) = 0$. The evolution of the perturbation $\boldsymbol{\epsilon}$ near the equilibrium $\textbf{x}^*$ is then given by:
\begin{equation}
\dot{\boldsymbol{\epsilon}}=\left(\textbf{Df}(\textbf{x}^*)+\frac{1}{2!}\boldsymbol{\epsilon}^T\textbf{D}^2\textbf{f}(\textbf{x}^*)\right)\boldsymbol{\epsilon}+\dots\, ,
\label{eq:taylor}
\end{equation}
where $\textbf{Df}(\textbf{x}^*)$ and $\textbf{D}^2\textbf{f}(\textbf{x}^*)$ are, respectively, the Jacobian matrix and the Hessian tensor~\footnote{With a slight abuse of language we denoted with such symbol the term $\sum_{ij}\partial_{ij}f_l(\textbf{x}^*)\epsilon_i\epsilon_j$ for each component $f_l$ of the vector $\textbf{f}=(f_1,\dots,f_n)$.} of the non-linear function $\textbf{f}$, both evaluated at the fixed point. Keeping only the first term on the right hand side corresponds to performing a linear analysis, i.e. computing the spectrum of $\textbf{Df}(\textbf{x}^*)$, and straightforwardly yields the definition of the dispersion relation, as introduced above.

Accounting for the second term in the expansion results in a quadratic problem that cannot be solved exactly. We can, however, make some progress by replacing $\boldsymbol{\epsilon}^T\textbf{D}^2\textbf{f}(\textbf{x}^*)/2$ with a constant factor that depends on the equilibrium point and also on the initial perturbation, $\boldsymbol{\epsilon}_0=\textbf{x}(0)-\textbf{x}^*$. In this way Eq.~\eqref{eq:taylor} can be written again as a linear system
\begin{equation}
\dot{\boldsymbol{\epsilon}}=\left[\textbf{Df}(\textbf{x}^*)+\frac{1}{2!}\boldsymbol{\epsilon}_0^T\textbf{D}^2\textbf{f}(\textbf{x}^*)\right]\boldsymbol{\epsilon}\equiv \left[\textbf{J}(\textbf{x}^*)+ \textbf{P}(\textbf{x}^*,\boldsymbol{\epsilon}_0)\right]\boldsymbol{\epsilon}\, ,
\label{eq:taylor2}
\end{equation}
where $\textbf{J}$ is the Jacobian and $\textbf{P}$ stands for a perturbation that is small, but not negligible. The spectrum of $\textbf{J}+\textbf{P}$ can thus provide information about the possible growth of the initial perturbation. In analogy with the  dispersion relation, and borrowing the name from the pseudospectrum, we define the {\em pseudo-dispersion relation} $\lambda_{\epsilon}$ to be the largest real part of the eigenvalues of $\textbf{J}+\textbf{P}$. A positive value for $\lambda_{\epsilon}$ implies that $\boldsymbol{\epsilon}$  grows in time. The perturbation hence is amplified and the system can possibly head towards a different attractor.

\begin{figure*}[t]\vspace*{-2.cm}
	\centering
	\includegraphics[scale=0.38]{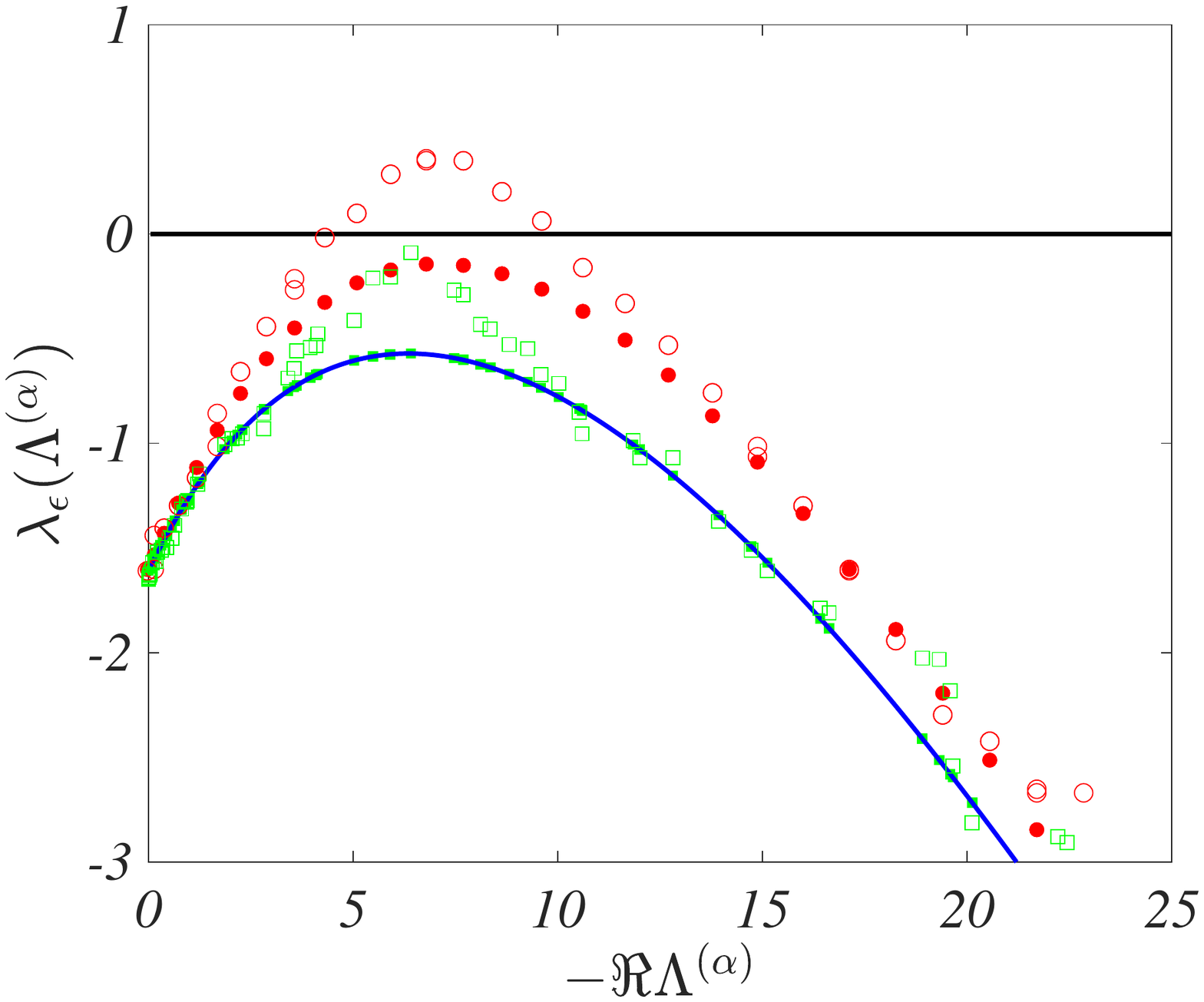}\quad \includegraphics[scale=0.38]{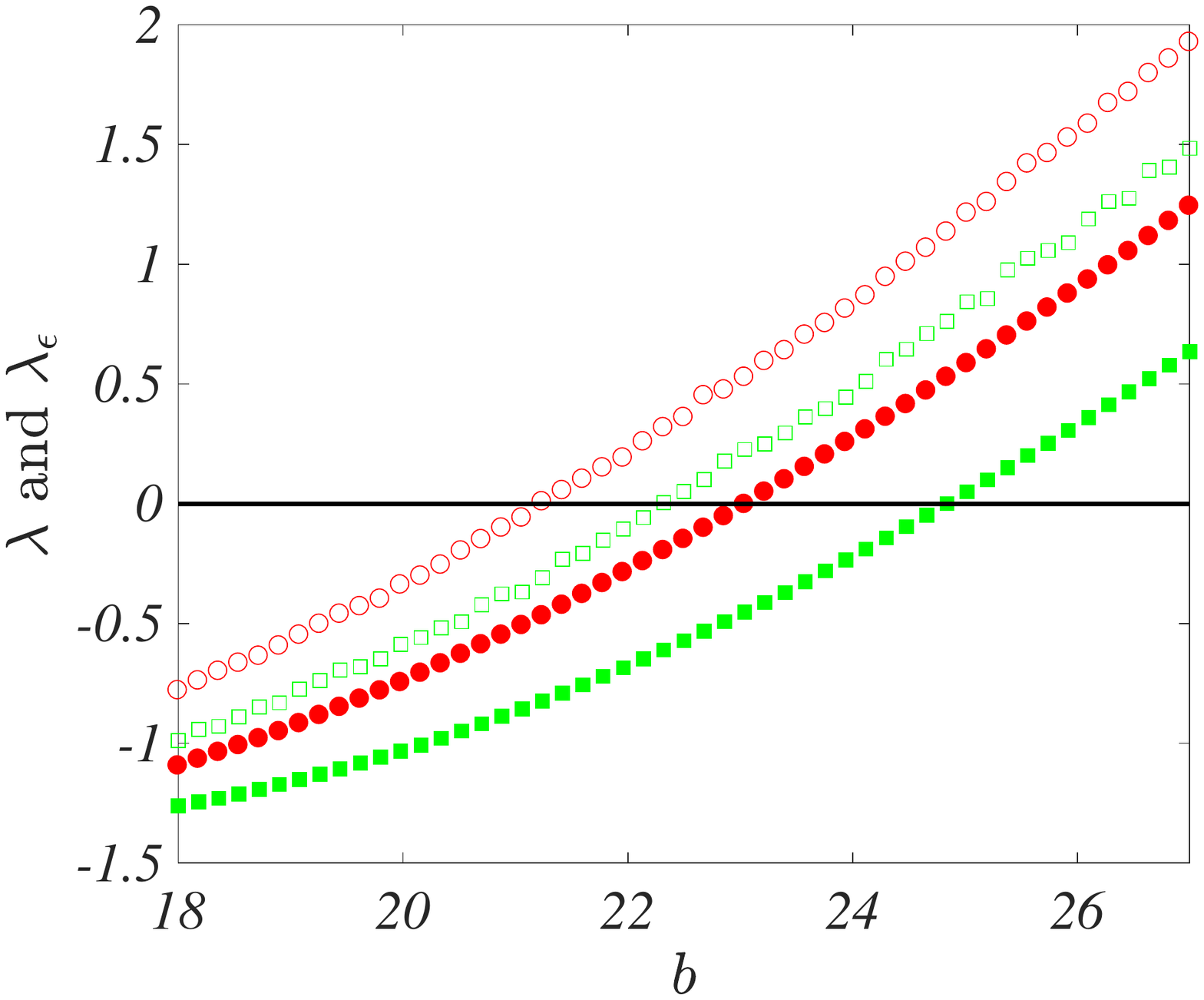}
	\vspace{-2cm}
	\caption{\textbf{Dispersion relation and pseudo-dispersion relation}. Left panel: Fixing the parameters $(b,c)=(22.5,61.1)$ (see also Fig.~\ref{fig:RelDispPatt}) we draw the dispersion relation for the non-normal (directed) network (red dots) and  symmetrised network (green dots). Empty symbols stand for the pseudo-dispersion relation, with an identical code for the assigned colours.  The system cannot exhibit Turing patterns because the dispersion relations are consistently negative. However the pseudo-dispersion relation is positive when dealing with a non-normal network. This  implies that the system undergoes an initial amplification which can eventually form a macroscopic pattern. {Right panel: using the same symbol/colour scheme used in the left panel, we show the maximum of the dispersion relation as a function of $b$. We can see that the pseudo-dispersion relation for the non-normal network (empty red circles) is positive for a much larger parameter range. Hence, non-normality of the embedding support favours the emergence of patterns.} Here $\delta=0.2$ {and each point is the average over $50$ independent replicas}. The remaining parameters, as well as the networks selection, are those used in Fig.~\ref{fig:NNPatt}.}
	\label{fig:psedusp}
\end{figure*}

Let us observe that the spectrum of $\textbf{J}+\textbf{P}$ is not, in general, a small perturbation of the spectrum of $\textbf{J}$. We cannot therefore use the tools developed in~\cite{multiplex,hatanakao} but rely instead on a continuation method (see Appendix~\ref{sec:Contmet}). Let us denote by $\eta$ the norm of $\textbf{P}$ and observe that if $\boldsymbol{\epsilon}_0\rightarrow 0$ then $\eta\rightarrow 0$. Let us denote by $\mu_\eta$ one eigenvalue of $\textbf{J}+\textbf{P}$, once $||\textbf{P}||=\eta$. Notice that, in doing this, we are abusing notation because $\textbf{P}$ - and thus its norm - depends on $\boldsymbol{\epsilon}_0$, so different initial perturbations can produce different $\textbf{P}$, but all with the same norm. We can, however, account for this fact by averaging $\mu_\eta$ over several initial conditions that share the condition $||\textbf{P}||=\eta$. For $\eta=0$, we have that $\mu_0$ is an eigenvalue of $\textbf{J}$ and thus we can find an eigenvalue $\Lambda^{(\alpha)}$ of the Laplacian matrix such that $\mu_0=\mu_0(\Lambda^{(\alpha)})$. Then varying continuously $s\in(0,\eta)$ we can follow, with continuity, $\mu_s(\Lambda^{(\alpha)})$. In Fig.~\ref{fig:psedusp} {(left panel)} we show $\Re\mu_s(\Lambda^{(\alpha)})$, for $s=0$ and $s=\eta$, as a function of $\Lambda^{(\alpha)}$ computed with this procedure and for the parameter values corresponding to the setting of the right panel of Fig.~\ref{fig:RelDispPatt}. We recall that for this parameter setting, the system defined on a  non-normal network exhibits patterns of non-normality, while it does not when confined on its symmetrised homologue. Remarkably, the pseudo-dispersion relation captures this difference, the latter being positive for the non-normal network (empty red circles) while it is negative for the symmetric network (empty green squares). {In the right panel of Fig.~\ref{fig:psedusp} we show the dispersion relation and the pseudo-dispersion relation as a function of the model parameter $b$ while fixing all the remaining ones.} The procedure here outlined can be hence regarded as a viable approach for predicting the stability of the system beyond the conventional linear order of approximation. 

\section{Conclusions}

The aim of this paper was to discuss a generalised route to  pattern formation. To this end we studied a reaction-diffusion system defined on a network and considered the interesting setting where the network is non-normal. The concept of non-normality is quite intriguing: a multidimensional linear system, ruled by a non-normal matrix, can display a short time amplification for the norm of an injected perturbation, also in the case when the latter is predicted to fade away asymptotically, because of the stability assumption. For the case at hand, the non-normality of the embedding support cooperates with the non-normality inherited by the reaction component, to trigger the system unstable, even if the corresponding dispersion relation is negative. Working with the Brusselator model, for illustrative purpose for its recognized pedagogical value, we proved that non-normality yields a contraction for the basin of attraction of the homogeneous fixed point, and consequently acts as the material drive for the onset of the instability.  This phenomenon can be quantitatively analysed based on the pseudo-dispersion relation, an effective tool for predicting the fate of the perturbation beyond  conventional linear analysis and accounting for non-linear corrections. 
Taken together, making the network of couplings non-normal favours {the spontaneous drive to macroscopic organization}, an observation that could potentially help in overcoming the limitations intrinsic to other paradigmatic {approaches to pattern formation. Non-normality could result in a novel route to }the emergence of self-organised structures, {widespread} patterns of complexity {of broad applied and fundamental relevance}~\cite{pastorsatorrasvespignani2010}. These conclusions are even more relevant given the recent account on the ubiquity of non-normality across different fields of investigations~\cite{nonnormalnet}. To conclude, we remark that we have here solely focused on a purely deterministic setting. It is tempting to speculate that non normality in a stochastic framework (the noise source being exogenous or endogenous), will eventually yield a more 
robust and flexible route to pattern formation in dynamics modeling~\cite{Tommaso,linampl,loopamplif}.

\appendix
\section{Continuation method}
\label{sec:Contmet}
The computation of the eigenvalues/eigenvectors of $\textbf{J}+\textbf{P}$ cannot be tackled with a perturbative scheme because, as previously mentioned, the  correction on $\textbf{J}$ is not sufficiently small.  A possible way to overcome this problem is to consider the modified matrix
\begin{equation}
\textbf{J}(s):=\textbf{J}+s\textbf{P}\, ,
\label{eq:JsH}
\end{equation}
where $s\in[0,1]$. Hence $\textbf{J}(s)$ interpolates between the extreme cases $\textbf{J}$, the known system, and $\textbf{J}+\textbf{P}$ the system under scrutiny. For all $s$ we are interested in finding the eigenvalues, hereby cast into a diagonal matrix $\textbf{D}(s)$, and left/right eigenvectors, again organised into a matrix form, $\textbf{U}(s)$ and $\textbf{V}(s)$, where we explicitly emphasise their dependence on $s$. In summary, we want to solve
\begin{equation}
\textbf{J}(s)\textbf{V}(s)=\textbf{V}(s)\textbf{D}(s)\text{ and }\textbf{U}(s)\textbf{J}(s)=\textbf{D}(s)\textbf{U}(s)\, .
\label{eq:JsH2}
\end{equation}

Let us now assume $s$ to be a variable and differentiate the previous equations with respect to $s$. Under the requirement of regularity we obtain for $\textbf{V}(s)$ (and similarly for $\textbf{U}(s)$):
\begin{equation}
\textbf{P}\textbf{V}(s)+\textbf{J}(s)\frac{d}{ds}\textbf{V}(s)=\left(\frac{d}{ds}\textbf{V}(s)\right)\textbf{D}(s)+\textbf{V}(s)\frac{d}{ds}\textbf{D}(s)\, .
\label{eq:JsH3a}
\end{equation}
Left multiply by $\textbf{U}(s)$ to obtain 
\begin{equation}
\textbf{U}\textbf{P}\textbf{V}+\textbf{D}\textbf{U}\frac{d}{ds}\textbf{V}=\textbf{U}\left(\frac{d}{ds}\textbf{V}\right)\textbf{D}+\textbf{U}\textbf{V}\frac{d}{ds}\textbf{D}\, ,
\label{eq:JsH3}
\end{equation}
where, to simplify notation, we removed the explicit dependence on $s$. 

This relation, and the other obtained for $\textbf{U}$, are matrix differential equations involving the unknown matrix functions $\textbf{U}$, $\textbf{V}$ and $\textbf{D}$. Considering the diagonal terms of Eq.~\eqref{eq:JsH3}, and recalling that $\textbf{D}$ is a diagonal matrix, we obtain~\footnote{We here assume that the notation of repeated indexes stand for a sum, that is $\textbf{U}_{il}\textbf{V}_{li}=\sum_l \textbf{U}_{il}\textbf{V}_{li}$.}:
\begin{equation}
(\textbf{U}\textbf{P}\textbf{V})_{ii}+\textbf{D}_{ii}\textbf{U}_{il}\frac{d}{ds}\textbf{V}_{li}=\textbf{U}_{il}\left(\frac{d}{ds}\textbf{V}_{li}\right)\textbf{D}_{ii}+\textbf{U}_{il}\textbf{V}_{lm}\frac{d}{ds}\textbf{D}_{mi}\, ,
\label{eq:JsH4}
\end{equation}
that is, the term involving the derivative of $\textbf{V}$ cancels out and we end up with the derivative of $\textbf{D}$:
\begin{equation}
\textbf{U}_{il}\textbf{V}_{li}\frac{d}{ds}\textbf{D}_{ii}=(\textbf{U}\textbf{P}\textbf{V})_{ii}\, ,
\label{eq:JsH5}
\end{equation}
that is 
\begin{equation}
\frac{d}{ds}\mu_{i}(s)=\frac{(\textbf{U}\textbf{P}\textbf{V})_{ii}}{(\textbf{U}\textbf{V})_{ii}}\, ,
\label{eq:JsH6}
\end{equation}
where we reintroduced the eigenvalues $\mu_i(s)=\textbf{D}_{ii}$.

Considering now the generic component $ij$ of Eq.~\eqref{eq:JsH3}, we can obtain, after some straightforward computation,
\begin{equation}
\textbf{U}_{il}\frac{d}{ds}\textbf{V}_{li}=-\frac{(\textbf{U}\textbf{P}\textbf{V})_{ij}}{\mu_i(s)-\mu_j(s)}+\frac{(\textbf{U}\textbf{V})_{ij}}{\mu_i(s)-\mu_j(s)}\frac{(\textbf{U}\textbf{P}\textbf{V})_{jj}}{(\textbf{U}\textbf{V})_{jj}}\, ,
\label{eq:JsH7}
\end{equation}
and
\begin{equation}
\left(\frac{d}{ds}\textbf{U}_{il}\right)\textbf{V}_{li}=\frac{(\textbf{U}\textbf{P}\textbf{V})_{ij}}{\mu_i(s)-\mu_j(s)}-\frac{(\textbf{U}\textbf{V})_{ij}}{\mu_i(s)-\mu_j(s)}\frac{(\textbf{U}\textbf{P}\textbf{V})_{ii}}{(\textbf{U}\textbf{V})_{ii}}\, .
\label{eq:JsH8}
\end{equation}

Under the assumption of non-degenerate eigenvalues the above matrix ODE can be numerically solved starting from the initial conditions $\mu_i(0)$ (eigenvalues of $\textbf{J}$), $\textbf{U}_0$ and $\textbf{V}_0$ (resp. left and right eigenvectors of $\textbf{J}$), up to $s=1$ providing in this way the required continuation of the eigenvalues. Proceeding along this line {and using a first order method to solve the ODE} returns the pseudo-dispersion relations displayed in Fig.~\ref{fig:psedusp}.

\section*{Acknowledgements} 
R.M. would like to thank the Erasmus+ program, Universit\`a  di Firenze and Universit\'e de Namur for funding his master internship in the group of Professor T.C.


\end{document}